\documentclass[superscriptaddress,aps,pra,twocolumn,showpacs,preprintnumbers,amsmath,amssymb,floatfix,groupedaddress]{revtex4}

\usepackage[dvips]{graphicx}
\usepackage{dcolumn}
\usepackage{bm}
\usepackage{amssymb}
\usepackage{xr}

\newcommand{\eop}{\mathcal{E}}

\begin{document}

\title{Photon storage in $\Lambda$-type optically dense atomic media. I. Cavity model}

\author{Alexey V. Gorshkov}
\author{Axel Andr\'e}
\author{Mikhail D. Lukin}
\affiliation{Physics Department, Harvard University, Cambridge, Massachusetts 02138, USA}
\author{Anders S. S{\o}rensen}
\affiliation{QUANTOP, Danish National Research Foundation Centre of
Quantum Optics, Niels Bohr Institute, DK-2100 Copenhagen {\O},
Denmark}

\date{\today}


\begin{abstract}
In a recent paper [Gorshkov \textit{et al.}, Phys. Rev. Lett. \textbf{98}, 123601 (2007)], we used a universal physical picture to optimize and demonstrate equivalence between a wide range of techniques for storage and retrieval of photon wave packets in $\Lambda$-type atomic media in free space, including the adiabatic reduction of the photon group velocity, pulse-propagation control via off-resonant Raman techniques, and photon-echo-based techniques. In the present paper, we perform the same analysis for the cavity model. In particular, we show that the retrieval efficiency is equal to $C/(1+C)$ independent of the retrieval technique, where $C$ is the cooperativity parameter. We also derive the optimal strategy for storage and, in particular, demonstrate that at any detuning one can store, with the optimal efficiency of $C/(1+C)$, any smooth input mode satisfying $T C \gamma \gg 1$ and a certain class of resonant input modes satisfying $T C \gamma \sim 1$, where $T$ is the duration of the input mode and $2 \gamma$ is the transition linewidth. In the two subsequent papers of the series, we present the full analysis of the free-space model and discuss the effects of inhomogeneous broadening on photon storage.
\end{abstract} 



\maketitle

\section{Introduction}

The faithful storage of a traveling light pulse in an atomic memory and the subsequent retrieval of the state are currently being pursued in a number of laboratories around the world \cite{kimble02, rempe02, polzik04, hau01, phillips01, eisaman05, kuzmich05, kimble05, vuletic06, hemmer01, hemmer02, manson05, manson06, afzelius06, kroll05}. A strong motivation for this research comes from the field of quantum communication, where quantum information is  easily transmitted by photons, but the photonic states need to be stored locally to process the information. Such  applications as well as other ideas from quantum-information science have led to a strong interest in techniques to facilitate a  controlled interaction between atoms and single photons \cite{bouwmeester00, DLCZ01}. A conceptually simple realization of a matter-light quantum interface consists of a single atom absorbing a single photon. However, due to the very weak coupling of a single atom to light, this approach is extremely challenging and requires the use of very high-finesse cavities to effectively increase the coupling \cite{kimble02, rempe02}. To circumvent the problem of weak coupling, it has recently been realized that one can use an optically thick ensemble of atoms, and several different proposals have been made for how external classical control fields can be used to controllably map photon states onto collective atomic states  \cite{lukin00fleischhauer00, kozhekin00, moiseev01, polzik04}. The goal in all of these approaches is to map an incoming signal pulse into a long-lived atomic coherence (referred to as a spin wave), so that it can be
later retrieved ``on demand" with the highest possible efficiency. Remarkable experimental progress has already been made toward the implementation of these protocols in atomic gases \cite{polzik04,hau01,phillips01,eisaman05,kuzmich05,kimble05,vuletic06} and in impurities embedded in a solid state material \cite{hemmer01,hemmer02,manson05,manson06, afzelius06}. 
A central question that emerges from these advances is which approach represents the best possible strategy for given experimental parameters and for desired memory characteristics, and how the control fields or possibly the shape of the input photon wave packet can be chosen to achieve the maximum efficiency. In a recent paper \cite{gorshkov07}, we presented a novel physical picture that unifies a wide range of different approaches to photon storage in $\Lambda$-type atomic media and yields the optimal control strategy. This picture is based on two key observations. First, we showed that the retrieval efficiency of any given stored spin wave depends only on the optical depth of the medium and not on the properties of the control pulse. Physically, this follows from the fact that the branching ratio between collectively enhanced emission into desired modes and spontaneous decay depends only on the optical depth. The second observation is that the optimal storage process is the time reverse of retrieval (see also \cite{moiseev01,kraus06,photonecho}). This universal picture implies that the maximum efficiency for the combined process of storage followed by retrieval is the same for all approaches considered and depends only on the optical depth \cite{polziknote}. The optimum can be attained by 
adjusting the control or the shape of the photon wave packet. In the present paper and in the two papers that follow, Refs.~\cite{paperII,paperIII}, which we will refer to henceforth as paper II and paper III, respectively, we present all the details behind this universal picture and the optimal control shaping that it implies, 
as well as consider several extensions of this analysis beyond the results of Ref.~\cite{gorshkov07}. In particular, in the present paper we discuss the cavity model to be compared in paper II to the free-space model. In paper II, the full analysis of the free-space model is presented, and, in addition, the effects of spin-wave decay and of nondegeneracy of the two lower levels of the $\Lambda$ system are discussed. Finally, in paper III, we generalize our treatment to two different regimes of inhomogeneous broadening: with and without redistribution between frequency classes during the storage time. 

\begin{figure}[htb]
\includegraphics[scale = 0.5]{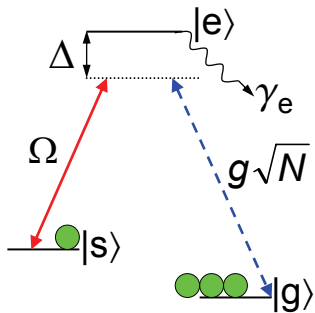}
\caption{(Color online) $\Lambda$-type medium coupled to a classical field (solid) with Rabi frequency $\Omega(t)$ and a quantum field (dashed). Due to collective enhancement \cite{lukin03}, the quantum field couples to the spin-wave excitations with an effective coupling constant $g \sqrt{N}$, where $N$ is the number of atoms in the medium. \label{fig:lambda}}
\end{figure}

A generic model for a quantum memory uses the $\Lambda$-type level configuration shown in
Fig.~\ref{fig:lambda}, in which a weak (quantum) signal field (the dashed line)
is detuned by a frequency $\Delta$ from the $|g\rangle-|e\rangle$ transition, whose optical coherence decays with rate $\gamma \geq \gamma_e/2$, where $\gamma_e$ is the spontaneous emission rate from state $|e\rangle$. A  copropagating
(classical) control beam (the solid line) with the same detuning $\Delta$ from the $|s\rangle-|e\rangle$ transition and time-dependent Rabi frequency envelope $\Omega$ is used to coherently manipulate the signal propagation and
map the photonic state onto the atoms, and vice versa.
In the present paper and in papers II and III, we discuss several different  approaches to photon storage, including far-off-resonant Raman, electromagnetically induced transparency (EIT), and photon-echo techniques. If we neglect the decay of the $|s\rangle-|g\rangle$ coherence, i.e., the decay of the spin wave, the only sources of loss in all of these approaches are the decay $\gamma$ of the optical polarization on the $|g\rangle-|e\rangle$ transition during both storage and retrieval, and the leakage of the pulse through the medium during storage. To achieve the maximum storage efficiency, one has to minimize  these two types of loss, and, in fact, as we will show in the present paper and in papers II and III, one has to make a compromise between them.

Higher optical depth increases the coherent coupling between the quantum signal field and the atoms and, thus, allows for higher photon storage efficiencies. It has therefore been suggested to put a cavity around the atomic ensemble \cite{DLCZ01, roch97, kuzmich98}, which will effectively increase the free-space optical depth $d$ by a factor equal to the number of passes a photon makes in the cavity before leaking out. We will denote this increased effective optical depth by the cooperativity parameter $C$. High-efficiency retrieval of a photon from an ensemble enclosed in a cavity has been recently demonstrated \cite{vuletic06}. In addition to being a promising experimental setup in itself \cite{vuleticnote}, the slightly simpler theoretical treatment of the cavity model offers a very useful tool for understanding the more complicated free-space model. Thus, in the present paper, we will treat photon storage in the cavity model, to be compared in paper II to the free-space model.

We will now review the three photon storage protocols (Raman, electromanetically induced tranparency, and photon echo) that are often discussed in the literature on photon storage and that we treat as special cases of our general formalism. The remainder of this section is intended as an introduction to both the present paper and paper II and will thus make use of the figures of merit of both the cavity model (the cooperativity parameter $C$) and the free-space model (the optical depth $d$). It will be implied in the following discussion that all the formulas containing $C$ ($d$) refer to the cavity (free-space) model. 

One possible strategy for light storage uses the Raman configuration, where the fields have a large detuning   (we will show that the appropriate limit is $|\Delta| \gg \gamma d$ or $|\Delta| \gg \gamma C$ rather than $|\Delta| \gg \gamma$, as one might naively assume by analogy with the single-atom case) and the photons are absorbed into the stable ground state $|s\rangle$ by stimulated Raman transitions \cite{kozhekin00,nunn06,mishina06}. 
With far-off-resonant interactions, the excited state $|e\rangle$ can be adiabatically eliminated to give simplified and solvable equations \cite{raymer81,raymer85}. Based on these simplified equations, the Raman scheme for storage of quantum states of light in atomic ensembles was proposed in Ref.~\cite{kozhekin00} and, simultaneously with the present work, has been analyzed in detail and optimized under the constraint of limited control power \cite{nunn06}. We show in the present paper and in paper II that, in the limit of large cooperativity parameter $C$ or large optical depth $d$, one can ignore the decay $\gamma$ of the optical $|g\rangle-|e\rangle$ coherence, as is done in Ref.~\cite{nunn06}.

An alternative storage strategy is based on electromagnetically induced transparency \cite{lukin00fleischhauer00, fleischhauer02, lukin03},  where resonant control fields  ($|\Delta| \ll \gamma d$ or $|\Delta| \ll \gamma C$) are used to open a spectral transparency window for the quantum field. In this approach, the quantum field travels at a reduced group velocity, which is then adiabatically reduced to zero. Similarly to the Raman case, the excited state can also be eliminated on resonance, provided the control field is sufficiently weak. This again simplifies the equations  to analytically solvable ones \cite{fleischhauer02}. 

We will treat both the far-off-resonant Raman scheme and the resonant EIT scheme as special cases of a more general ``adiabatic" limit, i.e., the limit in which the excited state can be adiabatically eliminated. We will show that, for the purposes of optimal photon storage, the condition of validity of the adiabatic elimination is almost independent of the single-photon detuning $\Delta$ (in particular, it is similar in the  Raman and resonant cases) and is given by  $T d \gamma \gg 1$ or $T C \gamma \gg 1$, where $T$ is the duration of the incoming pulse. We will show that, provided a (smooth) incoming photon wave packet is long enough that this condition is satisfied ($T \gg 1/(d \gamma)$ or $T \gg 1/(C \gamma)$), it can be stored with the maximum possible efficiency, which depends only on the optical depth $d$ or the cooperativity parameter $C$ and not on the detuning $\Delta$ or the shape of the wave packet. In the case of the cavity model discussed in the present paper, this maximum efficiency is simply given by $C/(1+C)$.

Finally, in the photon-echo-based approach to storage, a resonant photon is first allowed to get absorbed by the ensemble with the control field off. While the phrase ``photon echo" often refers to a wide class of experiments, we shall here consider a special case where one applies then a short resonant $\pi$ pulse, which maps excitations from the unstable excited state $|e\rangle$ into the stable ground state $|s\rangle$. Because this approach uses very short control pulses and because, as we will show, it is most efficient in storing short input pulses ($T \sim 1/(d \gamma)$ or $T \sim 1/(C \gamma)$), we will refer to this approach as ``fast" storage.  This technique was originally suggested in Ref.~\cite{moiseev01} for the case of Doppler-broadened atoms and has since been extensively studied both theoretically \cite{kraus06,kroll05,photonecho,paperIII} and experimentally \cite{manson06, afzelius06}. In Ref.~\cite{moiseev01}, it was noted that, if the photons are retrieved by using a control laser pulse traveling in the backward direction compared to storage, the Doppler broadening is reversed, and the dephasing occurring during storage is therefore also reversed, resulting in an ``echo," which may result in high efficiencies. In Refs.~\cite{kraus06,kroll05}, it was proposed to use controlled reversible inhomogeneous broadening (CRIB), that is, to artificially add an inhomogeneous broadening to an originally homogeneously broadened line and then to reverse this broadening to achieve an echo signal. 
In the present paper and in paper II, we consider a different limit of this proposal, where there is no inhomogeneous broadening of the optical transition, and storage is simply achieved by applying a fast $\pi$ pulse at the right time. Retrieval, which is accomplished with a second $\pi$ pulse, results \cite{scully06} in a directional output (as opposed to the loss due to the decay rate $\gamma $) exactly as in the adiabatic limit. We will show that in the limit of large $d$ or $C$ this procedure leads to an ideal storage and retrieval of the photonic state, while at every finite value of $d$ or $C$ there exists an optimal input photon mode that can be stored with efficiency equal to the maximum adiabatic storage efficiency (given by $C/(1+C)$ in the case of the cavity model). For comparison, in paper III, we will discuss how this approach measures up to the CRIB approach and show that adding and reversing inhomogeneous broadening as proposed in Refs.~\cite{kraus06,kroll05} may lead to an improvement in the storage efficiency, although the improvement is rather limited. 
   
The optimization of storage in all of these schemes consists of  finding the optimal balance between two sources of error: leakage of the input pulse through the ensemble and scattering of the input photons into $4 \pi$ due to spontaneous emission. In the EIT approach, a stronger control field is desirable, since it produces more robust interference and a wider transparency window, thus minimizing spontaneous emission losses. On the other hand, higher control power means larger group velocity and hence the inability to localize the input pulse inside the medium. The optimization in this case finds the optimal power and shape for the control field, given the duration and shape of the input pulse. In contrast, in the Raman scheme, a high value of $\Omega$ is required to have a sufficient coupling of the input photon to the spin wave \cite{nunn06}. On the other hand, large $\Omega$ will increase the decay rate due to spontaneous emission, which is  given by the optical pumping rate $\gamma \Omega^2/\Delta^2$. The optimization with respect to the shape and power of $\Omega$ for a given input mode again balances between these two sources of error. Finally, in the fast storage scheme the  control fields are fixed  to be perfect $\pi$ pulses, but one can  optimize with respect to the duration $T$ and the shape of the input mode. The input mode should be made as short as possible in order to avoid the loss due to optical polarization decay $\exp(-\gamma T)$. However, a mode that is too short will be too wide in frequency space and will not be absorbed by the ensemble (i.e., it will leak through). The optimization with respect to the duration and shape of the input mode finds the optimal balance between these two sources of error.

In all the photon storage techniques considered, ideal performance (i.e., unit efficiency) can be achieved in the limit of infinite optical depth $d$ or infinite cooperativity parameter $C$. For example, in the EIT regime in free space, the width of the spectral transparency window is  $\Delta \omega_\textrm{EIT} = v_\textrm{g} \sqrt{d}/L$, where $L$ is the length of the ensemble and $v_\textrm{g} \propto |\Omega|^2/d$ is the EIT group velocity \cite{fleischhauer05}. Thus, for a given $T$ and a given large value of $d$, one can first make $\Omega$, and hence $v_\textrm{g}$, small enough for the pulse to fit inside the medium. Then the enhancement of $\Delta \omega_\textrm{EIT}$ by an extra factor of $\sqrt{d}$ will ensure, if $d$ is sufficiently large, that the transparency window is still wide enough to induce negligible spontaneous emission. In the Raman regime, to avoid spontaneous emission decay via the optical pumping rate $\gamma \Omega^2/\Delta^2$, one should make $\Omega$ sufficiently small. If the optical depth or the cooperativity parameter is large enough, the coupling of the input mode to the atoms will then still be sufficient to avoid leakage even at this small value of $\Omega$. Finally, in fast storage, the pulse that is short enough ($T \gamma \ll 1$) to avoid optical polarization decay can still be absorbed in a free-space medium provided $d$ is large enough ($T d \gamma \gg 1$, as we will show in paper II). In the cavity model discussed in the present paper, due to the availability of only one spin-wave mode (the one that couples to the cavity mode), high-efficiency fast storage is harder to achieve than in free space: only pulses of a particular shape and duration ($T \sim 1/(C \gamma)$) give high fast storage efficiencies. 

Although ideal performance can be achieved at infinite optical depth, in practice, optical depth is always limited by experimental imperfections such as a limited number of atoms in a trap (e.g., Ref.~\cite{kuzmich05}, where the optical depth is roughly 8), competing four-wave mixing processes (e.g., Ref.~\cite{eisaman05}, where the optical depth is roughly 4), inhomogeneous broadening of impurity levels in solid state samples \cite{kroll05}, or other types of experimental imperfections. Therefore, the optimization of storage protocols at finite optical depth is essential.

Before proceeding with our analysis, we would like to specially note the recent work of Dantan \textit{et al.}, which also considers and illuminates some of the issues we discuss in the present paper  \cite{dantan04,dantan05a,dantan05b,dantan06} and in paper II \cite{dantan05b}. In particular, in Refs.~\cite{dantan04, dantan05a,dantan06}, focusing on broadband squeezed states as the input, the authors consider adiabatic storage in a cavity, derive an efficiency expression equivalent to ours, and recognize the interesting similarity between Raman and resonant regimes, both of which feature reduced sensitivity to spontaneous emission. We show in the present paper how, for the case of a single incoming spatiotemporal field mode, proper control field shaping can be used to achieve the same optimal efficiency independent of detuning. This effectively makes the Raman, the resonant, and the intermediate regimes all exhibit equally reduced sensitivity to spontaneous emission. In Ref. \cite{dantan05b}, as in paper II, the authors compare adiabatic storage in a cavity to adiabatic storage in free space and recognize important similarities and differences. In particular, it is recognized that the error in the cavity case scales as the inverse of the atomic density (as we also find in the present paper), while in free space it may scale as the inverse of the square root of the density. We show in paper II how proper optimization can be used to make the error in the free-space case also scale as the inverse of atomic density. We also bridge in free space the gap between the EIT and Raman cases, showing how proper control field shaping can be used to achieve the same optimal efficiency independent of detuning.

We would also like to note that the connection between optimal photon storage and time reversal that we present was first made in the context of photon-echo-based techniques. In particular, it was first shown in Ref.~\cite{moiseev01} and then discussed in detail in Refs.~\cite{kraus06, photonecho} that under certain conditions, such as high optical depth and sufficiently slow optical polarization decay rate, photon-echo techniques can result in ideal storage and retrieval, and that the retrieved photon field is then the time reverse of the original input field. We generalize this result in Ref.~\cite{gorshkov07}, in the present paper, and in papers II and III by demonstrating that the ideas of time reversal can be used to optimize photon storage even when the dynamics of the system are not completely reversible and when the ideal unit efficiency cannot be achieved. This is the case for finite cooperativity parameter $C$ in the cavity model and for finite optical depth $d$ in the free-space model. We also generalize the time-reversal-based optimization of photon storage from photon-echo-based techniques to any storage technique including, but not limited to, EIT and Raman techniques in homogeneously (present paper, Ref.~\cite{gorshkov07}, and paper II) and inhomogeneously (paper III) broadened $\Lambda$-type media.

We would also like to point out that mathematically some of the optimization problems we are solving in the present work (including Ref.\ \cite{gorshkov07}, the present paper, and papers II and III) fall into a rich and well-developed field of mathematics called optimal control theory \cite{pontryagin86,krotov96,bryson75}. In particular, we are interested in shaping the control pulse to maximize the storage efficiency (or efficiency of storage followed by retrieval) for a given input photon mode in the presence of optical polarization decay. Since in the equations of motion the control multiplies a dependent variable (optical polarization), this problem is a nonlinear optimal control problem \cite{pontryagin86}. A general solution to all nonlinear optimal control problems does not exist, so that such problems have to be treated on a case by case basis. We believe that the methods we suggest, such as the iterative time-reversal method introduced in Secs.~IV and V of paper II, may be useful in solving optimal control problems in other open (as well as closed) quantum systems. Similar iterative methods are a standard tool in applied optimal control \cite{krotov96,bryson75,krotov83, konnov99} and have been used for a variety of applications, including laser control of chemical reactions \cite{shapiro03, kosloff89}, design of NMR pulse sequences \cite{khaneja05}, loading of Bose-Einstein condensates into an optical lattice \cite{sklarz02}, and atom transport in time-dependent superlattices \cite{calarco04}. In fact, an optimization procedure that is based on gradient ascent \cite{khaneja05}, and that is very similar to that of Refs.\ \cite{krotov96,bryson75,krotov83, konnov99, shapiro03, kosloff89, khaneja05, sklarz02}, is directly applicable to our problem of finding the optimal control pulse, as we will discuss elsewhere \cite{gorshkov07b}. However, in the present paper and in papers II and III, we use time-reversal iterations for optimal control in a way different from the methods of Refs.\ \cite{krotov96,bryson75,krotov83, konnov99, shapiro03, kosloff89, khaneja05, sklarz02,calarco04}, as we will discuss in Sec.\ V of paper II. In particular, we will show that, in addition to being a convenient computational tool, our iterative optimization algorithm is, in fact, experimentally realizable \cite{novikova07}.

The remainder of this paper is organized as follows. Section \ref{sec:effic} applies to both the cavity and the free-space models and discusses our figure of merit for the performance of the photon storage. The rest of the paper discusses storage and retrieval of photons using homogeneously broadened atomic ensembles enclosed in a cavity. In Sec.~\ref{sec:cavmod}, we introduce the model. In Sec.~\ref{sec:cavopt}, without fully solving the equations analytically, we show that both the retrieval efficiency and the optimal storage efficiency are equal to $C/(1+C)$ (where $C$ is the cooperativity parameter), and derive the optimal storage strategy. In Secs.~\ref{sec:cavad} and \ref{sec:cavfast}, we solve the equations analytically in the adiabatic and fast limits, respectively, and demonstrate that the optimal storage efficiency can be achieved for any smooth input mode at any detuning satisfying $T C \gamma \gg 1$ and a certain class of resonant input modes satisfying $T C \gamma \sim 1$, where $T$ is the duration of the input mode. In Sec.~\ref{sec:cavsum}, we summarize the discussion of the cavity model. Finally, in the Appendixes, we present some details omitted in the main text. 

\section{Figure of merit \label{sec:effic}}

When comparing different storage and retrieval approaches, it is essential to have a figure of merit characterizing the performance of the memory. The discussion in this section of the appropriate figure of merit applies both to the cavity model discussed in this paper and to the free-space models discussed in papers II and III. Throughout this work we shall assume that we initially have a single incoming photon in a known spatiotemporal mode denoted by $\eop_\textrm{in}(t)$ (or, for the case of computing retrieval efficiency alone, a single excitation in a known atomic spin-wave mode).  We define the efficiency $\eta$ of all the mappings we consider (storage alone, retrieval alone, or storage followed by retrieval) as the probability to find the excitation in the output mode (photonic or atomic, as appropriate) after the interaction. Depending on the application one has in mind, this single-photon efficiency may or may not be the right quantity to consider, but provided that we are interested in a situation where we are mapping a single input mode into a single output mode, any other quantities may be derived from the single-photon efficiency $\eta$.

For all the interactions we consider, the full evolution results in a passive (beam-splitter-like) transformation 
\begin{equation}
\hat{b}_j=\sum_k U_{jk} \hat{a}_k,
\end{equation}
where $\hat{a}_j$ and $\hat{b}_k$ denote the annihilation operators for all the input and output modes, respectively (all photonic, spin-wave, and Langevin noise operators), with commutation relations $[\hat a_j, \hat a^\dagger_k] = \delta_{j,k}$ and $[\hat b_j, \hat b^\dagger_k] = \delta_{j,k}$. Here the matrix $U$ has to be unitary to preserve the commutation relations. The mapping from a certain input mode $\hat a_0$ to an output mode $\hat b_0$ with efficiency $\eta$ is therefore described by  $\hat b_0 = \sqrt{\eta} \hat a_0 + \sqrt{1-\eta} \hat c$, where $\hat c$ satisfies $[\hat c, \hat c^\dagger] = 1$ and represents some linear combination of all other input modes orthogonal to $\hat a_0$. If all input modes other than $\hat a_0$ are in the vacuum state, the parameter $\eta$ completely characterizes the mapping. If, for instance, the mode we are storing is in an entangled state with some other system $(|0\rangle_{\hat a_0} |x\rangle + |1\rangle_{\hat a_0} |y\rangle)/2$, where $|0\rangle_{\hat a_0}$ and  $|1\rangle_{\hat a_0}$ are the zero- and one-photon Fock states of the input mode, and  $|x\rangle$ and $|y\rangle$ are  two orthonormal states of the other system,  the fidelity of the entangled state after the mapping is easily found to be $F=(1+\eta)/2$. Similarly, Refs.~\cite{dantan04, dantan05a, dantan05b,dantan06} characterize the performance in terms of squeezing preservation parameter $\eta_\textrm{squeeze}$. If the input state is a squeezed vacuum state in a given mode $\hat a_0$, the squeezing preservation parameter can be shown to be equivalent to single-photon efficiency, i.e., $\eta_\textrm{squeeze} = \eta$. We will show below in the description of our model why in most experimental situations it is indeed reasonable to assume that the incoming noise (which is included in $\hat c$) is vacuum noise.

\section{Model \label{sec:cavmod}}

The details of the model and the derivation of the equations of motion are provided in Appendix \ref{sec:appModel}. In this section, we only give a brief introduction to the model and present the equations of motion without derivation.

We consider a  medium of $N$ $\Lambda$-type atoms with two metastable lower states, as shown in Fig.~\ref{fig:lambda}, interacting with two single-mode fields.  We neglect reabsorption of spontaneously emitted photons and treat the problem in a one-dimensional approximation. The $|g\rangle-|e\rangle$ transition  of frequency $\omega_{eg}$ of each of the atoms is coupled to a quantized traveling-wave cavity radiation mode (e.g., a mode of a ring cavity with one of the mirrors partially transmitting) with frequency $\omega_1 = \omega_{eg} - \Delta$ described by a slowly varying annihilation operator $\eop(t)$. The cavity decay rate is $2 \kappa$ and the corresponding input-output relation is \cite{walls94}
\begin{equation} \label{io}
\hat \eop_\textrm{out}(t) = \sqrt{2 \kappa} \hat \eop(t) - \hat \eop_\textrm{in}(t).
\end{equation}
In addition, the transitions $|s\rangle-|e\rangle$ of frequency $\omega_{es}$ are driven by a single-mode copropagating classical plane-wave control field 
 with frequency $\omega_2 = \omega_{es}-\Delta$ (i.e., at two-photon resonance $\omega_1-\omega_2=\omega_{sg}$, where $\hbar \omega_{sg}$ is the energy  difference between the two metastable states) described by a slowly varying Rabi frequency envelope $\Omega(t)$. 

In the dipole and rotating-wave approximations, assuming that almost all atoms are in the ground state at all times, and defining the polarization annihilation operator $\hat P(t) = \hat \sigma_{ge}(t)/\sqrt{N}$ and the spin-wave annihilation operator $\hat S(t) = \hat \sigma_{gs}(t)/\sqrt{N}$ (where $\hat \sigma_{\mu \nu}$ are slowly varying collective atomic operators defined in Appendix \ref{sec:appModel}), to first order in $\hat \eop$, the Heisenberg equations of motion are
\begin{eqnarray}\label{caveqs1}
\dot {\hat \eop} &=& -\kappa \hat \eop + i g \sqrt{N} \hat P + \sqrt{2 \kappa} \hat \eop_\textrm{in},
\\ \label{caveqs2}
\dot {\hat P} &=& - (\gamma+i \Delta) \hat P + i g \sqrt{N} \hat \eop + i \Omega \hat S + \sqrt{2 \gamma} \hat F_P,
\\ \label{caveqs3}
\dot {\hat S} &=& -\gamma_\textrm{s} \hat S + i \Omega^* \hat P + \sqrt{2 \gamma_s} \hat F_S,
\end{eqnarray}
where we have introduced the polarization decay rate $\gamma$, the spin-wave decay rate $\gamma_\textrm{s}$, and the corresponding Langevin noise operators $\hat F_P$ and $\hat F_S$. The coupling constant $g$ (assumed to be real for simplicity) between the atoms and the quantized field mode is collectively enhanced \cite{lukin03} by a factor of $\sqrt{N}$ to $g\sqrt{N}$.

As described in Appendix \ref{sec:appModel}, under reasonable experimental conditions, the incoming noise described by $\hat F_P$ and $\hat F_S$ is vacuum, i.e., all normally ordered noise correlations are zero. This is precisely the reason why, as noted in Sec.~\ref{sec:effic}, efficiency is the only number we need in order to fully characterize the mapping. 

We assume that all atoms are initially pumped into the ground state, i.e., no $\hat P$ or $\hat S$ excitations are present in the atoms. We also assume that the only input field excitations initially present are in the quantum field mode with an envelope shape $h_0(t)$ nonzero on $[0,T]$. The goal is to store the state of this mode in $\hat S$ and, starting at a time $T_\textrm{r} > T$, retrieve it back into a field mode. Since we are interested only in computing efficiencies (defined below) and since the incoming noise is vacuum, we can ignore the noise operators in Eqs.~(\ref{caveqs1})-(\ref{caveqs3}) and treat these equations as complex number equations. During storage, the initial conditions are $P(0) = 0$, $S(0) = 0$, and the input mode is $\eop_\textrm{in}(t) = h_0(t)$ (normalized according to $\int_0^T d t |\eop_\textrm{in}(t)|^2 = 1$). We have here dropped the carets on the operators to denote their complex number representations. The storage efficiency is then
\begin{equation}
\eta_{\textrm{s}} = \frac{(\textrm{number of stored excitations})}{(\textrm{number of incoming photons})} = |S(T)|^2.
\end{equation}
Similarly, during retrieval, the initial and boundary conditions are $P(T_\textrm{r}) = 0$, $S(T_\textrm{r}) = S(T)$, and $\eop_\textrm{in}(t) = 0$. $\eop_\textrm{out}(t)$ then represents the shape of the quantum mode into which we retrieve, and the total efficiency of storage followed by retrieval is given by
\begin{equation} \label{etatot}
\eta_{\textrm{tot}} = \frac{(\textrm{number of retrieved photons})}{(\textrm{number of incoming photons})}=\int_{T_\textrm{r}}^\infty \! d t |\eop_\textrm{out}(t)|^2.
\end{equation}
If we instead take $S(T_\textrm{r}) = 1$, we obtain the retrieval efficiency: 
\begin{equation} \label{etar}
\eta_{\textrm{r}} = \frac{(\textrm{number of retrieved photons})}{(\textrm{number of stored excitations})}
\!=\! \int_{T_\textrm{r}}^\infty d t |\eop_\textrm{out}(t)|^2. 
\end{equation}

From now on we will neglect the slow decay of the spin wave (i.e., set $\gamma_\textrm{s} = 0$) but, as briefly discussed below at the ends of Secs.~\ref{sec:cavadret} and \ref{sec:cavadst}, spin-wave decay is not hard to include. Nonzero $\gamma_\textrm{s}$ will simply introduce an exponential decay without making the solution or the optimal control shaping harder.

To get the closest analogy to the free-space regime, we assume we are always in the ``bad cavity" limit ($\kappa \gg g \sqrt{N}$), in which $\eop$ in Eq.~(\ref{caveqs1}) can be adiabatically eliminated to give
\begin{eqnarray} \label{caveout}
\eop_\textrm{out} &=& \eop_\textrm{in} + i \sqrt{2 \gamma C} P,
\\ \label{cavPeq}
\dot P &=& - (\gamma (1+C)+i \Delta) P + i \Omega S + i \sqrt{2 \gamma C} \eop_\textrm{in}, 
\\ \label{cavSeq}
\dot S &=& i \Omega^* P, 
\end{eqnarray} 
where $C= g^2 N/(\kappa \gamma)$ is the cooperativity parameter. To relate to the free space situation discussed in paper II, we can write the cooperativity parameter as $C = 2 d \left[(1/(2 \kappa))/(L/c)\right]$, where $d = g^2 N L/(c \gamma)$ is the definition of optical depth  used in the free-space model of paper II and where the factor in the square brackets (proportional to cavity finesse) is equal to the number of passes a photon would make through an empty cavity before leaking out (i.e., the photon lifetime in the cavity divided by the time a single pass takes). Thus, up to a factor of order unity, the cooperativity parameter $C$ represents the effective optical depth of the medium in the cavity, so that the efficiency dependence on $C$ in the cavity should be compared to the efficiency dependence on $d$ in free space. We note that, although Eqs.~(\ref{caveout})-(\ref{cavSeq}) describe our case of quantized light coupled to the $|g\rangle-|e\rangle$ transition, they will also precisely be the equations describing the propagation of a classical probe pulse. To see this one can simply take the expectation values of Eqs.~(\ref{caveqs1})-(\ref{caveqs3}) and use the fact that classical probe pulses are described by coherent states.

It is convenient to reduce Eqs.~(\ref{cavPeq}) and (\ref{cavSeq}) to a single equation 
\begin{equation} \label{cavoneeq}
\left[ \ddot S - \frac{\dot \Omega^*}{\Omega^*} \dot S\right] + (\gamma (1+C) + i \Delta) \dot S + |\Omega|^2 S = 
-\Omega^* \sqrt{2 \gamma C} \eop_\textrm{in}.
\end{equation}
This second-order differential equation cannot, in general, be fully solved analytically. However, in the next section we derive a number of exact results about the optimal efficiency anyway. 

\section{Optimal Strategy for Storage and Retrieval \label{sec:cavopt}}

In this section, we derive several important results regarding the optimal strategy for maximizing the storage efficiency, the retrieval efficiency, and the combined (storage followed by retrieval) efficiency without making any more approximations.

It is convenient to first consider retrieval. Although we cannot, in general, analytically solve for the output field $\eop_\textrm{out}(t)$, we will now show that the retrieval efficiency is always $C/(1+C)$ independent of the detuning $\Delta$ and the control shape $\Omega(t)$ provided that no excitations are left in the atoms at $t = \infty$, i.e., $P(\infty) = 0$ and $S(\infty) = 0$. From Eqs.~(\ref{cavPeq}) and (\ref{cavSeq}) and using $\eop_\textrm{in}(t) = 0$, we find 
\begin{equation} \label{cavcons}
\frac{d}{d t}\left( |P|^2 + |S|^2\right) = - 2 \gamma (1+C) |P|^2. 
\end{equation}
Using this and Eqs.~(\ref{etar}) and (\ref{caveout}), the retrieval efficiency becomes 
\begin{equation} \label{cavetar}
\eta_{\textrm{r}} = \frac{C}{1+C} \left( |S(T_\textrm{r})|^2 + |P(T_\textrm{r})|^2 - |S(\infty)|^2 - |P(\infty)|^2 \right),
\end{equation}
which reduces to $C/(1+C)$ for $S(T_\textrm{r})=1$, $P(T_\textrm{r})=P(\infty)=S(\infty)=0$. The value of the retrieval error ($1-\eta_\textrm{r} = 1/(1+C)$) and its independence from $\Delta$ and $\Omega$ follow directly from the branching ratio between the decay rates in Eq.~(\ref{cavPeq}) (or equivalently in Eq.~(\ref{cavcons})). The decay rate for $P$ into undesired modes is $\gamma$, while the decay rate for $P$ into the desired mode $\eop_\textrm{out}$ is $\gamma C$. The retrieval efficiency, which is the ratio between the desired decay rate and the total decay rate, is, therefore, equal to $C/(1+C)$ independent of the control field.

We have thus shown that, provided our control pulse is sufficiently long and/or powerful to leave no excitations in the atoms (we will refer to this as complete retrieval), the retrieval efficiency is independent of $\Delta$ and $\Omega(t)$ and is always equal to $C/(1+C)$. Therefore, any control field is optimal for  retrieval provided it pumps all excitations out of the system. Using this knowledge of the retrieval efficiency, in the remainder of this section we will use a time-reversal argument to deduce the optimal storage strategy and the optimal storage efficiency. Here we will only give the essence of and the intuition behind the time-reversal argument, and leave the derivation to Secs.~IV and V of paper II. In the remainder of the paper, we will independently confirm the validity of this argument in the adiabatic and fast limits. 
 
Applied to the present situation, the essence of the time-reversal argument is as follows. Suppose one fixes the cooperativity parameter $C$ and the detuning $\Delta$ and considers complete retrieval from the spin wave with a given control field $\Omega(t)$ into an output mode $\eop_\textrm{out}(t)$ of duration $T_\textrm{out}$. According to the time-reversal argument, the efficiency for storing the time reverse of the output field ($\eop_\textrm{in}(t) = \eop^*_\textrm{out}(T_\textrm{out}-t)$) with $\Omega^*(T_\textrm{out}-t)$, the time reverse of the retrieval control field, into the spin wave is equal to the retrieval efficiency \cite{complexconjugate}. Although this claim is not trivial to prove (see paper II), it is rather intuitive: since the retrieval procedure can be regarded as a generalized beam-splitter-like transformation (Sec.~\ref{sec:effic}), the equality of the two efficiencies is simply the statement that the probability of going from a given input port of the beam splitter to a given output port is equal to the probability of going backward from that output port to the original input port.

Therefore, the time-reversal argument shows that the maximum efficiencies for storage and storage followed by retrieval are $C/(1+C)$ (i.e., the retrieval efficiency) and $C^2/(1+C)^2$ (i.e., its square), respectively. Moreover, it says that these maximum efficiencies are obtained if the input field $\eop_\textrm{in}(t)$ and the storage control field $\Omega(t)$ are such that $\Omega^*(T-t)$, i.e., the time reverse of $\Omega(t)$, retrieves the spin-wave excitation into the output mode $\eop_\textrm{out}(t)=\eop^*_\textrm{in}(T-t)$, i.e., the time reverse of $\eop_\textrm{in}(t)$. In order to say for which input fields the optimal storage control $\Omega(t)$ can be found (or, equivalently, into which output fields a spin-wave excitation can be retrieved), we need to consider the limits, in which Eq.~(\ref{cavoneeq}) can be fully solved analytically. These limits, adiabatic and fast, will be discussed in the following sections. 

\section{Adiabatic Retrieval and Storage\label{sec:cavad}}

\subsection{Adiabatic retrieval \label{sec:cavadret}}

In the previous section, we have found, based on time reversal, the maximum storage efficiency and the scenario under which it can be achieved. Since the optimal storage into a given input mode requires the ability to carry out optimal retrieval into the time reverse of this mode,   we will, in the following sections (Secs.~\ref{sec:cavad} and \ref{sec:cavfast}), solve Eq.~(\ref{cavoneeq}) analytically in two important limits to find out which modes we can  retrieve into and store optimally. The first such limit, which we will discuss in this section (Sec.~\ref{sec:cavad}), corresponds to smooth control and input fields, such that the term in the square brackets in Eq.~(\ref{cavoneeq}) can be dropped. This ``adiabatic'' limit corresponds to an adiabatic elimination of $P$ in Eq.~(\ref{cavPeq}). The precise conditions for this approximation will be discussed in Sec.~\ref{sec:cavadcond}. In this section, we discuss the retrieval process. 

It is instructive to recognize that in the adiabatic approximation (i.e., with $\dot P$ in Eq.~(\ref{cavPeq}) replaced with $0$), if one uses rescaled variables $\eop_\textrm{in}(t)/\Omega(t)$, $\eop_\textrm{out}(t)/\Omega(t)$, and $P(t)/\Omega(t)$ and makes a change of variables $t \rightarrow h(T_\textrm{r},t)$, where 
\begin{equation} \label{hdefcav}
h(t,t') = \int_t^{t'} |\Omega(t'')|^2 dt'',
\end{equation} 
then Eqs.~(\ref{caveout})-(\ref{cavSeq}) become independent of $\Omega$ and can be solved in this $\Omega$-independent form, so that for any given $\Omega$ the solution in the original variables would follow by simple rescaling. However, since the equations are sufficiently simple and in order to avoid confusion introduced by additional notation, we will solve Eqs.~(\ref{caveout})-(\ref{cavSeq}) directly without making the change of variables. 

To compute the output field during adiabatic retrieval, we assume for simplicity that retrieval begins at time $t=0$ rather than at time $t=T_\textrm{r}$ and adiabatically eliminate $P$ in Eqs.~(\ref{cavPeq}) and (\ref{cavSeq}) (i.e., replace $\dot P$ in Eq.~(\ref{cavPeq}) with zero) to obtain a first-order linear ordinary differential equation for $S$. Then, using $S(0) = 1$ and $\eop_\textrm{in}(t) = 0$, we solve this equation to find 
\begin{equation} \label{cavadeout}
\eop_\textrm{out}(t) = - \sqrt{2 \gamma C} \frac{\Omega(t)}{\gamma (1+C) + i \Delta} e^{-\frac{1}{\gamma (1+C) + i \Delta} h(0,t)}. 
\end{equation}
The $t$-dependent phase $i h(0,t) \Delta /(\gamma^2 (1+C)^2 + \Delta^2)$ in the last factor is the ac Stark shift, which results in a shift of the output field frequency away from bare two-photon resonance. Computing the retrieval efficiency using Eq.~(\ref{cavadeout}), we find  
\begin{equation}
\eta_\textrm{r} = \frac{C}{1+C} \left(1 - e^{-\frac{2 \gamma (1+C)}{\gamma^2 (1+C)^2 + \Delta^2} h(0,\infty )}\right),
\label{etaCavfiniteh}  
\end{equation}
which is equal to $C/(1+C)$ provided the control pulse is sufficiently powerful and/or long to ensure  that
\begin{equation} \label{cavrethcond}
\frac{2 \gamma (1+C)}{\gamma^2 (1+C)^2 + \Delta^2} h(0,\infty ) \gg 1,
\end{equation}
which is the same as the condition $P(\infty) = S(\infty) = 0$. Note that adiabatic elimination did not affect the exact value of the efficiency and kept it independent of $\Omega(t)$ and $\Delta$ by preserving the branching ratio between the desired and undesired state transfers. Also note that,  unlike the general argument in the previous section, which assumed 
$P(\infty) = S(\infty) = 0$, Eq.~(\ref{etaCavfiniteh}) allows for the precise calculation of the retrieval efficiency for any $h(0,\infty)$.  

As noted in the Introduction, two important subsets of the adiabatic limit, the resonant limit and the Raman limit, are often discussed in the literature. Although, as we show in this work, the basic physics based on the branching ratio and time-reversal arguments is shared by both of these approaches to quantum memory, a more detailed discussion of the physics behind them involves significant differences. In fact, prior to this work, the fact that the two approaches are in a sense equivalent was not recognized to our knowledge: only interesting similarities were pointed out \cite{dantan04,dantan05a,dantan05b}. As an example of an important difference, the resonant and Raman limits give different dependences on $C$ of the duration of the output pulse in Eq.~(\ref{cavadeout}):
\begin{equation}
T_\textrm{out} \sim \frac{\gamma^2 C^2 + \Delta^2}{\gamma C |\Omega|^2},
\end{equation}
where we assumed $C \gtrsim 1$. In the resonant limit ($\gamma C \gg |\Delta|$), $T_\textrm{out} \sim \gamma C/|\Omega|^2$, while in the Raman limit ($\gamma C \ll |\Delta|$), $T_\textrm{out} \sim \Delta^2/(\gamma C |\Omega|^2)$. It is worth emphasizing that the Raman limit condition is $\gamma C \ll |\Delta|$ and not $\gamma \ll |\Delta|$, as one may naively think by analogy with the single-atom case.

It follows from the concept of time reversal
that the modes that  can be stored optimally are the time reverses of the modes onto which a spin wave can be retrieved. We will now show that, in the adiabatic limit, at any given $\Delta$ and $C$, we can shape $\Omega(t)$ to retrieve onto any normalized mode $e(t)$. Integrating the norm squared of Eq.~(\ref{cavadeout}) with $\eop_\textrm{out}(t) = \sqrt{C/(1+C)} e(t)$, we get
\begin{equation} \label{eeq}
\int_0^t d t' |e(t')|^2 = 1-e^{-\frac{2 \gamma (1+C) h(0,t)}{\gamma^2 (1+C)^2 + \Delta^2} }.
\end{equation}
Solving this equation for $h(0,t)$ and then taking the square root of its derivative with respect to $t$, we find $|\Omega(t)|$. Knowing $h(0,t)$, the phase of $\Omega(t)$ can be determined from Eq.~(\ref{cavadeout}). Putting the magnitude and the phase together, we have
\begin{equation} \label{cavretoptom}
\Omega(t) = - \frac{\gamma(1+C)+ i \Delta}{\sqrt{2 \gamma (1+C)}} \frac{e(t)}{\sqrt{\int_t^\infty  |e(t')|^2 d t'}} e^{-i \frac{\Delta h(0,t)}{\gamma^2 (1+C)^2 + \Delta^2}},
\end{equation}
where $h(0,t)$ should be determined from Eq.~(\ref{eeq}). For any $e(t)$, this expression gives the control $\Omega(t)$ that retrieves the spin wave into that mode. The phase of $\Omega(t)$, up to an unimportant constant phase, is given by the phase of the desired output mode plus compensation for the Stark shift (the last factor). It is also worth noting that, up to a minus sign and a factor equal to the first fraction in Eq.~(\ref{cavretoptom}), $\Omega(t)$ is simply equal to $e(t)/S(t)$. 

We note that, if one wants to shape the retrieval into a mode $e(t)$ that drops to zero at some time $T_\textrm{out}$ sufficiently rapidly, $|\Omega(t)|$ in Eq.~(\ref{cavretoptom}) will go to $\infty$ at $t = T_\textrm{out}$. The infinite part can, however, be truncated without significantly affecting the efficiency or the precision of $e(t)$ generation. One can confirm that the loss in efficiency is small by inserting into the adiabatic solution in Eq.~(\ref{etaCavfiniteh}) a value of $h(0,\infty)$ that is finite but large enough to satisfy Eq.~(\ref{cavrethcond}). One can similarly confirm that the generation of $e(t)$ can be precise with truncated control fields by using Eq.~(\ref{cavadeout}). However, to be completely certain that the truncation is harmless, one has to solve Eqs.~(\ref{caveout})-(\ref{cavSeq}) numerically without making the adiabatic approximation. We will do this in Sec.~\ref{sec:cavadcond} for the case of storage, where the same truncation issue is present. 

We briefly mention that the spin-wave decay rate $\gamma_\textrm{s}$, which we have ignored so far, simply introduces a decay described by $\exp(-\gamma_\textrm{s} t)$ into Eq.~(\ref{cavadeout}) and, unless we retrieve much faster than $1/\gamma_\textrm{s}$, makes retrieval efficiency control dependent. With nonzero $\gamma_\textrm{s}$, we can still shape retrieval to go into any mode: we shape the control using Eq.~(\ref{cavretoptom}) as if there were no $\gamma_\textrm{s}$ decay except that the desired output mode $e(t)$ should be replaced with the normalized version of $e(t) \exp(\gamma_\textrm{s} t)$, i.e.,
\begin{equation}
e(t) \rightarrow e(t) e^{\gamma_\textrm{s} t} \left[\int_0^\infty d t' |e(t')|^2 e^{2 \gamma_\textrm{s} t'}\right]^{-\frac{1}{2}}.
\end{equation}
The retrieval efficiency will, however, be output-mode-dependent in this case: it will be multiplied (and hence reduced) by $\left[\int_0^\infty d t' |e(t')|^2 \exp(2 \gamma_\textrm{s} t')\right]^{-1}$.

\subsection{Adiabatic storage \label{sec:cavadst}}

In principle, using the solution for retrieval from the previous section, the time-reversal argument of Sec.~IV immediately guarantees that, provided we are in the adiabatic limit (conditions to be discussed in Sec.~\ref{sec:cavadcond}), we can always shape the control field to store any input mode $\eop_\textrm{in}(t)$ at any detuning $\Delta$ with the maximum efficiency $C/(1+C)$. However, for completeness, and to verify that the optimal storage control field is indeed the time reverse of the control field that retrieves into $\eop^*_\textrm{in}(T-t)$, we give in this section the solution to adiabatic storage.   

In the adiabatic approximation, we use a procedure very similar to that used in the retrieval solution, to find
\begin{equation} \label{cavstor}
S(T) = \sqrt{\frac{C}{1+C}} \int_0^T dt f(t) \eop_\textrm{in}(t),  
\end{equation}
where
\begin{equation} \label{cavadf}
f(t) = -\frac{\Omega^*(t) \sqrt{2 \gamma (1+C)}}{\gamma(1+C) + i \Delta} e^{-\frac{h(t,T)}{\gamma(1+C)+i \Delta}}.
\end{equation}
The storage efficiency is then
\begin{equation} \label{cavetas}
\eta_\textrm{s} = \frac{C}{1+C} \left|\int_0^T dt f(t) \eop_\textrm{in}(t)\right|^2.
\end{equation}
We are interested in computing the control that maximizes $\eta_\textrm{s}$ for a given $\eop_\textrm{in}(t)$. We find in Appendix \ref{sec:appstor} that the maximum storage efficiency is $C/(1+C)$ and that it can be achieved (in the adiabatic limit) for any $\Delta$ and $\eop_\textrm{in}(t)$, and that the optimal control is
\begin{equation} \label{cavstoptom}
\Omega(t) = - \frac{\gamma(1+C)- i \Delta}{\sqrt{2 \gamma (1+C)}} \frac{\eop_\textrm{in}(t)}{\sqrt{\int_0^t  |\eop_\textrm{in}(t')|^2 d t'}} e^{i \frac{\Delta h(t,T)}{\gamma^2 (1+C)^2 + \Delta^2}},
\end{equation}
where $h(t,T)$ can be found by inserting Eq.~(\ref{cavstoptom}) into Eq.~(\ref{hdefcav}). The phase of $\Omega(t)$, up to an unimportant constant phase, is thus given by the phase of the input mode plus compensation for the Stark shift (the last factor). As for the retrieval control discussed in the previous section, we note that, although $|\Omega(t)|$ in Eq.~(\ref{cavstoptom}) goes to $\infty$ at $t=0$, the infinite part can be truncated without significantly affecting the efficiency. This can be confirmed analytically using Eq.~(\ref{cavetas}) provided the adiabatic limit is satisfied. We will also confirm this numerically in the next section without making the adiabatic approximation.

As expected from the time-reversal argument, the optimal control we derived is just the time reverse ($\Omega(t) \rightarrow \Omega^*(T-t)$) of the control that retrieves into $\eop^*_\textrm{in}(T-t)$, the time reverse of the input mode. We verify this in Appendix \ref{sec:appstor}.   

Although optimal storage efficiencies are the same in the Raman and adiabatic limits, as in the case of retrieval, rather different physical behavior can be seen in the two limits. It is now the dependence on $C$ of the optimal control intensity (which can be found from Eq.~(\ref{cavstoptom})) that can be used to separate resonant and Raman behavior. Assuming for simplicity $C \gtrsim 1$, in the resonant limit ($\gamma C \gg |\Delta|$), $|\Omega| \sim \sqrt{\gamma C/T}$, while in the Raman limit ($\gamma C \ll |\Delta|$), $|\Omega| \sim |\Delta|/\sqrt{\gamma C T}$. Since complete retrieval and optimal storage are just time reverses of each other, it is not surprising that these relations are identical to the ones we derived for the dependence of output pulse duration on $C$ in the previous section. This opposite dependence of $|\Omega|$ on $C$ in the Raman and EIT limits is, in fact, the signature of a simple physical fact: while the coupling of the input photon to the spin wave increases with increasing $\Omega$ in the Raman case, it effectively decreases in the EIT regime where a very large $\Omega$ will give a very wide transparency window and a group velocity equal to the speed of light. This is why as the cooperativity parameter changes, the control has to be adjusted differently in the two regimes.

As for retrieval, we briefly mention that nonzero $\gamma_\textrm{s}$ simply introduces $\exp(-\gamma_\textrm{s}(T-t))$ decay into Eq.~(\ref{cavadf}). The optimal storage control can still be found using Eq.~(\ref{cavstoptom}) as if there were no decay, except that the input mode should be replaced with the normalized version of $\eop_\textrm{in}(t) \exp(-\gamma_\textrm{s}(T-t))$, i.e.,
\begin{equation}
\eop_\textrm{in}(t) \rightarrow \eop_\textrm{in}(t) e^{-\gamma_\textrm{s}(T-t)} \left[\int_0^T   d t' |\eop_\textrm{in}(t')|^2 e^{-2 \gamma_\textrm{s}(T-t')}\right]^{-\frac{1}{2}}\!.
\end{equation}
However, the optimal storage efficiency will now depend on input pulse duration and shape: it will be multiplied (and hence reduced) by $\int_0^T d t' |\eop_\textrm{in}(t')|^2 \exp(-2 \gamma_\textrm{s}(T-t'))$. It is important to note that with nonzero spin-wave decay the optimal storage efficiency of a particular input mode is no longer identical to the retrieval efficiency into its time reverse. This is not at variance with the time-reversal argument discussed in detail in paper II (which still applies when $\gamma_\textrm{s} \neq 0$), since the corresponding optimal storage and retrieval control shapes are no longer the time reverses of each other, and, in contrast to the $\gamma_\textrm{s} = 0$ case, the retrieval efficiency is now control dependent. Finally, we note that, when we consider storage followed by retrieval, in order to take into account the spin-wave decay during the storage time $[T,T_\textrm{r}]$, one should just multiply the total efficiency by $\exp(-2 \gamma_\textrm{s} (T_\textrm{r}-T))$. 

\subsection{Adiabaticity conditions\label{sec:cavadcond}} 

We have found that, provided we are in the adiabatic limit, any input mode can be stored optimally. In this section we show that, independent of $\Delta$, the sufficient and necessary condition for optimal adiabatic storage of a pulse of duration $T$ to be consistent with the adiabatic approximation is $T C \gamma \gg 1$.
 
To find the conditions for the adiabatic elimination of $P$ in Eq.~(\ref{cavPeq}), we do the elimination and then require its consistency by enforcing \cite{cavinitialcondition} 
\begin{equation} \label{cavPdot}
|\dot P| \ll |(\gamma C+i \Delta) P|
\end{equation}
(we assume for simplicity throughout this section that $C \gtrsim 1$).
During retrieval, sufficient conditions for Eq.~(\ref{cavPdot}) are 
\begin{eqnarray} \label{cavcond1}
|\Omega| \ll |\gamma C+i \Delta|,
\\ \label{cavcond2}
\left|\frac{\dot \Omega}{\Omega}\right| \ll |\gamma C+i \Delta|,
\end{eqnarray}
which limit, respectively, the power and the bandwidth of the control pulse. These are easily satisfied in practice by using sufficiently weak and smooth retrieval control pulses.

During storage, the satisfaction of Eq.~(\ref{cavPdot}) requires, in addition to conditions (\ref{cavcond1}) and (\ref{cavcond2}), the satisfaction of
\begin{equation} \label{cavcond3}
\left|\frac{\dot \eop_\textrm{in}}{\eop_\textrm{in}}\right| \ll |\gamma C+i \Delta|,
\end{equation}
which limits the bandwidth of the input pulse. In particular, for a smooth input pulse of duration $T$, this condition is implied by
\begin{equation} \label{cavcond}
T C \gamma \gg 1.
\end{equation}
Let us now show that for optimal storage, the condition (\ref{cavcond}) also implies conditions (\ref{cavcond1}) and (\ref{cavcond2}) and is thus the only required adiabaticity condition (provided $\eop_\textrm{in}$ is smooth). Application of Eq.~(\ref{cavstoptom}) reduces Eq.~(\ref{cavcond1}) to Eq.~(\ref{cavcond}). Equation (\ref{cavcond2}), in turn, reduces to the conditions on how fast the magnitude $|\Omega|$ and the phase $\phi$ (which compensates for the Stark shift) of the control can change:
\begin{equation} \label{cavomchange}
\left|\frac{\frac{d}{dt} |\Omega|}{|\Omega|}\right|,\left|\dot \phi\right| \ll |\gamma C+i \Delta|,
\end{equation} 
where $\Omega = |\Omega| \exp(i \phi)$. Application of Eq.~(\ref{cavstoptom}) shows that Eq.~(\ref{cavcond}) implies Eq.~(\ref{cavomchange}). 

We have thus shown that $T C \gamma \gg 1$ is a sufficient condition for the validity of adiabatic elimination in optimal storage. But, in fact, from the amplitude of the optimal storage control field (Eq.~(\ref{cavstoptom})), one can see that Eq.~(\ref{cavcond1}) implies that $T C \gamma \gg 1$ is also a necessary condition for the validity of adiabatic elimination in optimal storage. (To show that $T C \gamma \gg 1$ is a necessary condition, one also has to use the extra condition on the adiabatic value of $P(0)$ \cite{cavinitialcondition} to rule out the special situation when $\Omega(t)$ is such that Eq.~(\ref{cavPdot}) is satisfied but Eq.~(\ref{cavcond1}) is not.)

\begin{figure}[ht]
\begin{center}
\includegraphics[scale = 1]{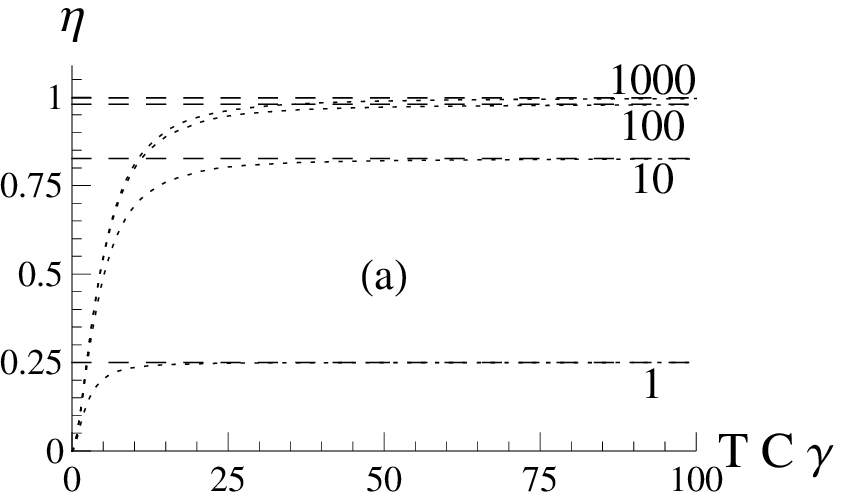}
\includegraphics[scale = 1]{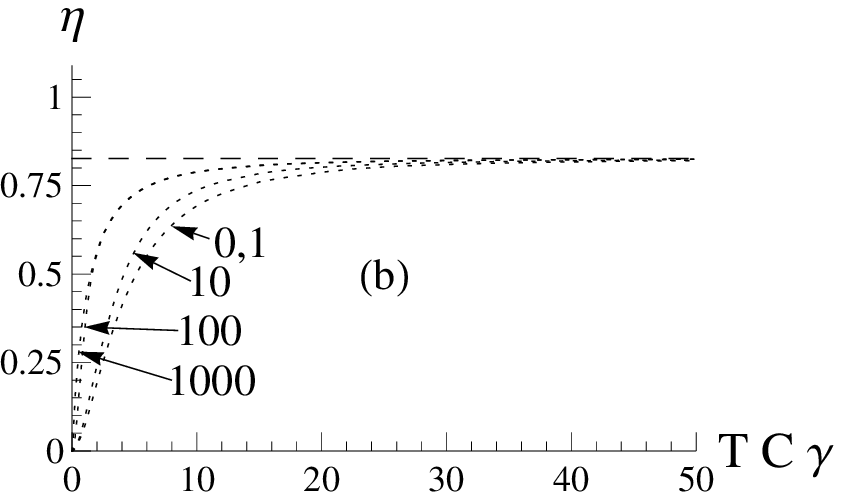}
\end{center}
\caption{Breakdown of optimal adiabatic storage in a cavity at $T C \gamma \lesssim 10$. In (a), the total efficiency of storage followed by retrieval is plotted for $\Delta = 0$ and $C = 1, 10, 100$, and $1000$. The horizontal dashed lines are the maximal values $(C/(1+C))^2$. Dotted lines are obtained for the input from Eq.~(\ref{Gaussiancav}) using adiabatic Eq.~(\ref{cavstoptom}) to shape the storage control but then using exact Eqs.~(\ref{cavPeq}) and (\ref{cavSeq}) to numerically compute the efficiency. In (b), the same plot is made for $C = 10$ and $\Delta/\gamma = 0, 1, 10, 100$, and $1000$.   \label{fig:cavadbreak}}
\end{figure}

To verify the adiabaticity condition in Eq.~(\ref{cavcond}) and investigate the breakdown of adiabaticity for short input pulses, we consider a Gaussian-like input mode (shown in Fig.~3 of paper II) 
\begin{equation} \label{Gaussiancav}
\eop_\textrm{in}(t) = A(e^{-30 (t/T - 0.5)^2} - e^{-7.5})/\sqrt{T},
\end{equation} 
where for computational convenience we have required $\eop_\textrm{in}(0) = \eop_\textrm{in}(T)=0$ and where $A \approx 2.09$ is a normalization constant. We fix the cooperativity parameter $C$, the detuning $\Delta$, and the pulse duration $T$, and use Eq.~(\ref{cavstoptom}) to shape the control field. We then use Eqs.~(\ref{cavPeq}) and (\ref{cavSeq}) \textit{without} the adiabatic approximation to calculate numerically the actual storage efficiency that this control field gives, and multiply it by the control-independent retrieval efficiency $C/(1+C)$, to get the total efficiency of storage followed by retrieval. As we decrease $T$, we expect this efficiency to fall below $(C/(1+C))^2$ once $T C \gamma \gg 1$ is no longer satisfied. And indeed in Fig.~\ref{fig:cavadbreak}(a) we observe this behavior for $\Delta = 0$ and $C = 1, 10, 100, 1000$. In Fig.~\ref{fig:cavadbreak}(b), we fix $C = 10$ and show how optimal adiabatic storage breaks down at different detunings $\Delta$ from $0$ to $1000 \gamma$. From Fig.~\ref{fig:cavadbreak}(b), we see that, as we move from the resonant limit ($C \gamma \gg |\Delta|$) to the Raman limit ($C \gamma \ll |\Delta|$), we can go to slightly smaller values of $T C \gamma$ before storage breaks down. However, since the curves for $\Delta = 100 \gamma$ and $\Delta = 1000 \gamma$ almost coincide, it is clear that $T C \gamma \gg 1$ is still the relevant condition no matter how large $\Delta$ is, which must be the case since the condition (\ref{cavcond1}) breaks down for shorter $T$. The most likely reason why in the Raman limit adiabaticity is slightly easier to satisfy is because in the Raman limit it is only condition (\ref{cavcond1}) that reduces to $T C \gamma \gg 1$, while conditions (\ref{cavcond2}) and (\ref{cavcond3}) reduce to $T \Delta \gg 1$, which is weaker than $T C \gamma \gg 1$ (since $\Delta \gg C \gamma$ in the Raman limit). In the resonant limit, in contrast, all three conditions (\ref{cavcond1})-(\ref{cavcond3}) reduce to $T C \gamma \gg 1$.   

Before turning to the discussion of fast retrieval and storage, we note that the use of Eq.~(\ref{cavstoptom}) to calculate the storage control fields for Fig.~\ref{fig:cavadbreak} resulted in a control field $\Omega(t)$ whose magnitude went to $\infty$ at $t = 0$, as predicted in the previous section. To generate Fig.~\ref{fig:cavadbreak}, the optimal $|\Omega(t)|$ were therefore cut off for $t < T/100$ to take the value $|\Omega(T/100)|$. The fact that the optimal efficiency of $(C/(1+C))^2$ represented by the dashed lines in Fig.~\ref{fig:cavadbreak} is still achieved by the dotted curves, despite the use of truncated controls, proves that truncation of the storage control does not significantly affect the storage efficiency. Since the retrieval field generation is directly related to optimal storage by time reversal, as explained in Sec.~\ref{sec:cavopt}, this also means that truncating retrieval controls does not significantly affect the precision with which a given retrieval mode $e(t)$ can be generated. The losses associated with truncation are insignificant only if the conditions in Eq.~(\ref{cavrethcond}) and Eq.~(\ref{cavsthcond}) are still satisfied for the truncated retrieval and storage control fields, respectively. If the limit on the control pulse energy is so tight that these conditions are not satisfied, a separate optimization problem, which is beyond the scope of the present paper, has to be solved.     

\section{Fast Retrieval and Storage \label{sec:cavfast}}

We have shown that adiabatic storage allows us to store optimally a mode of duration $T$, having any smooth shape and any detuning $\Delta$, provided that the adiabaticity condition $T C \gamma \gg 1$ is satisfied. In this section we solve Eq.~(\ref{cavoneeq}) analytically in the second important limit, the so-called ``fast" limit, and show that this limit allows for optimal storage of a certain class of input modes of duration $T \sim 1/(C \gamma)$.   

The fast limit corresponds to the situation when $\Omega$ is very large during a short control pulse ($|\Omega| \gg C \gamma$ and $|\Omega| \gg |\Delta|$), so that we can neglect all terms in  Eq.~(\ref{cavoneeq}) except $|\Omega|^2S$ and $\ddot{S}$. This corresponds to keeping only terms containing $\Omega$ on the right-hand side of Eqs.~(\ref{cavPeq}) and (\ref{cavSeq}) and results in undamped Rabi oscillations between optical and spin polarizations $P$ and $S$. One can use this limit to implement a ``fast'' storage scheme, in which the input pulse is resonant ($\Delta = 0$) and the control pulse is a short $\pi$ pulse at $t = T$, as well as fast retrieval, in which the control is a $\pi$-pulse at $t=T_\textrm{r}$. Provided the $\pi$ pulse is applied on resonance and at approximately constant intensity, the condition for perfect $\pi$ pulse performance is $|\Omega| \gg C \gamma$ (assuming $C \gtrsim 1$). To fully describe these processes, we furthermore need to solve Eq.~(\ref{cavoneeq}) while the control is off, which can also be done analytically.

During fast retrieval, assuming the $\pi$ pulse takes place at time $t = 0$ instead of time $t = T_\textrm{r}$ and assuming the $\pi$ pulse is perfect, the initial $S = 1$ results after the $\pi$ pulse in $P = i$. We then solve for $P(t)$ from Eq.~(\ref{cavPeq}) and insert the solution into Eq.~(\ref{caveout}) to obtain 
\begin{equation} \label{cavfasteout}
\eop_\textrm{out}(t) = - \sqrt{2 \gamma C} e^{- \gamma (1+C) t}.
\end{equation}
Consistent with the general expression in Eq.~(\ref{cavetar}) and the branching ratio argument in Sec.~\ref{sec:cavopt}, the retrieval efficiency is again $C/(1+C)$.

An alternative explanation for why the fast retrieval gives the same retrieval efficiency as the adiabatic retrieval is that, thanks to the adiabatic elimination of $P$, the adiabatic limit effectively describes a two-level system. Therefore, Eq.~(\ref{cavfasteout}) is in fact a special case of Eq.~(\ref{cavadeout}) with 
\begin{equation}\label{cavfastom}
\Omega(t) = (\gamma(1+C) + i \Delta) e^{-i \Delta t}.
\end{equation} 
Although, at this $\Omega(t)$, Eq.~(\ref{cavadeout}) is not a good approximation to the actual output field because, for example, condition (\ref{cavcond1}) is not satisfied, this illustrates the equivalence of the two approaches. 

Since the control field in fast retrieval is not adjustable (it is always a perfect $\pi$ pulse), fast retrieval gives only one possible output mode, that of Eq.~(\ref{cavfasteout}). By time reversal, the time reverse of this mode of duration $T \sim 1/(\gamma C)$ is thus the only mode that can be optimally stored (with efficiency $C/(1+C)$) using fast storage at this $C$.
  
For completeness and to confirm the time-reversal argument, the optimal input mode for fast storage can also be calculated directly. 
For an input mode $\eop_\textrm{in}(t)$ that comes in from $t = 0$ to $t=T$, assuming a perfect $\pi$ pulse at $t=T$, we find by a method similar to the one used in fast retrieval that
\begin{equation} \label{cavfastst}
S(T) = \sqrt{\frac{C}{1+C}} \int_0^T d t f(t) \eop_\textrm{in}(t),
\end{equation}
where
\begin{equation} \label{cavfastf}
f(t) = - e^{\gamma (1+C) (t-T)} \sqrt{2 (1+C) \gamma}.
\end{equation}
Similarly to retrieval, Eq.~(\ref{cavfastf}) is a special case of Eq.~(\ref{cavadf}) with $\Omega(t) = (\gamma (1+C)-i \Delta) \exp(i \Delta (T-t))$ (the time reverse of the right-hand side of Eq.~(\ref{cavfastom})). Since $f(t)$ is real and normalized according to $\int^T_0 f(t)^2 dt =1$, this integral is a scalar product similar to Eq.~(\ref{cavetas}) discussed in Appendix \ref{sec:appstor}, and the optimal fast storage efficiency of $C/(1+C)$ is achieved for a single input mode $\eop_\textrm{in}(t) = f(t)$ (up to an arbitrary overall unimportant phase). This optimal $\eop_\textrm{in}(t)$ is precisely the (renormalized) time reverse of the output of fast readout in Eq.~(\ref{cavfasteout}), as expected by time reversal. 

Two comments are in order regarding the optimal input pulse $\eop_\textrm{in}(t) = f(t)$.  
First, we would like to note that short exponentially varying pulses, as in our optimal solution $\eop_\textrm{in}(t) = f(t)$, have been proposed before to achieve efficient photon-echo-based storage \cite{kalachev}.
Second, it is worth noting that, although $\eop_\textrm{in}(t) = f(t)$ gives the optimal storage, generating such short exponentially rising pulses may in practice be hard for high $C$. Since the efficiency is given by the overlap of $\eop_\textrm{in}(t)$ with $f(t)$ (see Eq.~(\ref{cavfastst})), fast storage in a cavity is inferior in this respect to fast storage in free space, because in the latter case any input pulse satisfying $T \gamma \ll 1$ and $d T \gamma \gg 1$ results in storage efficiency close to unity \cite{paperII}. 

\section{Summary \label{sec:cavsum}}

In conclusion, we have treated in detail the storage and retrieval of photons in homogeneously broadened $\Lambda$-type atomic media enclosed in a running-wave cavity. We have shown that, provided that no excitations are left in the atoms at the end of the retrieval process, the retrieval efficiency is independent of the control and the detuning and is equal to $C/(1+C)$. We have also derived the optimal strategy for storage in the adiabatic and fast limits and, therefore, demonstrated that one can store, with the optimal efficiency of $C/(1+C)$, any smooth input mode satisfying $T C \gamma \gg 1$ and having any detuning $\Delta$ and a certain class of resonant input modes satisfying $T C \gamma \sim 1$. We have also noted that the optimal storage control field for a given input mode is the time reverse of the control field that accomplishes retrieval into the time reverse of this input mode. This fact and the equality of maximum storage efficiency and the retrieval efficiency are, in fact, the consequence of a general time-reversal argument to be presented in detail in paper II. In paper II, we will also present the full discussion of photon storage in homogeneously broadened $\Lambda$-type atomic media in free space, while in paper III, we will consider the effects of inhomogeneous broadening on photon storage.

Finally, it is important to note that, to achieve the optimal efficiencies derived in the present paper, it is necessary to have rigid temporal synchronization between the input pulse and the storage control pulse, which may become difficult in practice for short input pulses. In fact, since there is only one accessible atomic mode in the case of homogeneously broadened media enclosed in a cavity (unless one varies the angle between the control and the input \cite{vuletic06}), this temporal synchronization is necessary to obtain high efficiencies even if the cooperativity parameter is very large. This problem can, however, be alleviated whenever multiple atomic modes are accessible, which is the case for homogeneously broadened media in free space considered in paper II and for inhomogeneously broadened media considered in paper III. In those cases, infinite optical depth allows one to achieve unit efficiency without rigid synchronization. However, despite this disadvantage of the cavity setup, we will discuss in paper II that the cavity setup is superior to the free-space setup in other respects, such as the enhancement of the optical depth by the cavity finesse and the avoidance of the unfavorable scaling of the error as $1/\sqrt{N}$ (vs $1/C \propto 1/N$), which sometimes occurs in the free-space model.
 
\section{Acknowledgments}
We thank M.~Fleischhauer, M.~D.~Eisaman, E.~Polzik, J.~H.~M\"{u}ller, A.~Peng, I.~Novikova, D.~F.~Phillips, R.~L.~Walsworth, M.~Hohensee, M.~Klein, Y.~Xiao, N.\ Khaneja, A.~S.~Zibrov, P.~Walther, and A.~Nemiroski for fruitful discussions. This work was supported by the NSF, Danish Natural Science Research Council, DARPA, Harvard-MIT CUA, and Sloan and Packard Foundations.


\appendix
\section{Details of the Model and the Derivation of the Equations of Motion 
\label{sec:appModel}} 

In Sec.~\ref{sec:cavmod}, we gave a brief introduction to the model and presented the equations of motion without derivation. In this appendix, the details of the model and the derivation of the equations of motion (\ref{caveqs1})-(\ref{caveqs3}) are provided.

The electric-field vector operator for the cavity field is given by
\begin{equation}
\mathbf{\hat{E}}_1(z) = \mathbf{\epsilon}_1 \left(\frac{\hbar \omega_1}{2 \epsilon_0 V}\right)^{1/2} \left(\hat a e^{i \omega_1 z/c}+ \hat a^\dagger e^{-i \omega_1 z/c}\right),
\end{equation}
where $\hat a^\dagger$ is the mode creation operator, $\omega_1$ is the mode frequency, $\mathbf{\epsilon}_1$ is the polarization unit vector, $\epsilon_0$ is the permittivity of free space, $V$ is the quantization volume for the field, and $c$ is the speed of light.

The copropagating single-mode classical plane-wave control field with frequency $\omega_2$ is described by an electric-field vector  
\begin{equation}
\mathbf{E}_2(z,t) = \mathbf{\epsilon}_2 \eop_2(t) \cos(\omega_2(t-z/c)),
\end{equation}
where $\mathbf{\epsilon}_2$ is the polarization unit vector, and $\eop_2(t)$ is the amplitude. Then, using the dipole and rotating-wave approximations, the Hamiltonian is
\begin{eqnarray} \label{cavH1} 
\hat H \!&=& \hat H_0 + \hat V,
\\ \label{cavH2}  
\hat H_0 \!& = & \hbar \omega_1 \hat a^\dagger \hat a + \sum^N_{i=1} \left(\hbar \omega_{se} \hat \sigma^i_{ss} + \hbar \omega_{ge} \hat \sigma^i_{ee}\right),
\\ \label{cavH3}
\hat V \!& = & - \sum^N_{i=1} \mathbf{\hat d_i} \cdot (\mathbf{E}_2 (z_i,t)+ \mathbf{\hat E}_1(z_i)) 
\\ \nonumber 
\!& = & \!\! - \hbar\! \sum^N_{i=1}\! \left(\Omega (t) \hat \sigma^i_{es} e^{-i \omega_2 (t-z_i/c)}\!+\! \hat a g e^{i \omega_1 z_i/c} \hat \sigma^i_{eg}\!\right)\! +\! \textrm{h.c.}. 
\end{eqnarray}
Here $\textrm{h.c.}$ stands for Hermitian conjugate, $\hat \sigma^i_{\mu \nu} = |\mu\rangle_{i i}\langle\nu|$ is the internal state operator of the $i$th atom between states $|\mu\rangle$ and $|\nu \rangle$, $z_i$ is the position of the $i$th atom, $\mathbf{\hat d_i}$ is the dipole moment vector operator for the $i$th atom, $\Omega(t) = {_i\langle} e |(\mathbf{\hat d_i} \cdot \mathbf{\epsilon}_2)|s \rangle_i \eop_2 (t)/(2 \hbar)$ (assumed to be equal for all $i$) is the Rabi frequency of the classical field, and $g = {_i\langle} e |(\mathbf{\hat d_i} \cdot \mathbf{\epsilon}_1)|g \rangle_i \sqrt{\frac{\omega_1}{2 \hbar \epsilon_0 V}}$ (assumed to be real for simplicity and equal for all $i$) is the coupling constant between the atoms and the quantized field mode. We note that, in order to avoid carrying extra factors of $2$ around, $\Omega$ is defined as half of the traditional definition of the Rabi frequency, so that a $\pi$ pulse, for example, takes time $\pi/(2 \Omega)$.

In the Heisenberg picture, we introduce slowly varying collective atomic operators
\begin{eqnarray}
\hat \sigma_{\mu \mu} &=& \sum_i \hat \sigma_{\mu \mu}^{i},
\\
\hat \sigma_{es} &=& \sum_i \hat \sigma_{es}^{i} e^{-i \omega_2 (t-z_i/c)},
\\
\hat \sigma_{eg} &=& \sum_i \hat \sigma_{eg}^{i} e^{-i \omega_1 (t-z_i/c)},
\\
\hat \sigma_{sg} &=& \sum_i \hat \sigma_{sg}^{i} e^{-i (\omega_1-\omega_2) (t-z_i/c)},
\end{eqnarray}
and a slowly varying cavity mode annihilation operator
\begin{equation}
\hat \eop = \hat a e^{i \omega_1 t},
\end{equation}
which satisfy same-time commutation relations
\begin{eqnarray} \label{cavcomrel}
\left[\hat \sigma_{\mu \nu}(t), \hat \sigma_{\alpha \beta}(t)\right] &=& \delta_{\nu \alpha} \hat \sigma_{\mu \beta}(t) - \delta_{\mu \beta} \hat \sigma_{\alpha \nu}(t),
\\
\left[\hat \eop(t),\hat \eop^\dagger(t)\right] &=& 1,
\end{eqnarray}
and yield an effective rotating frame Hamiltonian
\begin{equation}
\hat{\tilde H} = \hbar \Delta \hat \sigma_{ee}- (\hbar \Omega (t) \hat \sigma_{es} + \hbar g \hat \eop \hat \sigma_{eg} + h.c.).
\end{equation}
The equations of motion are then given by
\begin{eqnarray} \label{fullcavityeq}
\dot {\hat \eop} \!\!&=&\!\! -\kappa \hat \eop + i g \hat \sigma_{ge} + \sqrt{2 \kappa} \hat \eop_\textrm{in},
\nonumber \\
\dot {\hat \sigma}_{gg} \!\!&=&\!\! \gamma_\textrm{eg} \hat \sigma_{ee} - i g \hat \eop \hat \sigma_{eg} + i g \hat \eop^\dagger \hat \sigma_{ge} + \hat F_{gg},
\nonumber \\
\dot {\hat \sigma}_{ss} \!\!&=&\!\! \gamma_\textrm{es} \hat \sigma_{ee} - i \Omega \hat \sigma_{es} + i \Omega^* \hat \sigma_{se} + \hat F_{ss},
\nonumber \\
\dot {\hat \sigma}_{ee} \!\!&=&\!\! - \gamma_\textrm{e} \hat \sigma_{ee} \!+ i \Omega \hat \sigma_{es} \!- i \Omega^* \hat \sigma_{se}\! + i g \hat \eop \hat \sigma_{eg}\! - i g \hat \eop^\dagger \hat \sigma_{ge} \!+\! \hat F_{ee},
\nonumber \\
\dot {\hat \sigma}_{ge} \!\!&=&\!\! -(\gamma+i \Delta) \hat \sigma_{ge} + i \Omega \hat \sigma_{gs} + i g \hat \eop (\hat \sigma_{gg} - \hat \sigma_{ee}) + \hat F_{ge},
\nonumber \\
\dot {\hat \sigma}_{es} \!\!&=&\!\! -(\gamma' - i \Delta) \hat \sigma_{es} + i \Omega^* (\hat \sigma_{ee}-\hat \sigma_{ss}) - i g \hat \eop^\dagger \hat \sigma_{gs} + \hat F_{es},
\nonumber \\
\dot {\hat \sigma}_{gs} \!\!&=&\!\! -\gamma_\textrm{s} \hat \sigma_{gs} + i \Omega^* \hat \sigma_{ge} - i g \hat \eop \hat \sigma_{es} + \hat F_{gs}, 
\end{eqnarray}
with the input-output relation for the quantum field given by Eq.~(\ref{io}). 

In Eqs.~(\ref{fullcavityeq}) we have introduced decay, which, in turn, necessitated the introduction of Langevin noise operators $\hat F_{\mu \nu}$ for the atomic operators and the input field $\hat \eop_\textrm{in}$ for the quantum field. The radiative decay rate of the excited state $|e \rangle$ is $\gamma_\textrm{e} = \gamma_\textrm{es} + \gamma_\textrm{eg}$, the sum of decay rates into $|s\rangle$ and into $|g\rangle$. The decay rate of optical coherence $\hat \sigma_{ge}$ is $\gamma = \gamma_\textrm{e}/2 + \gamma_\textrm{deph}$ where, in addition to radiative decay, we allow for extra dephasing, such as, for example, that caused by collisions with buffer gas atoms in warm vapor cells. Similarly, the decay rate $\gamma' = \gamma_\textrm{e}/2 + \gamma'_\textrm{deph}$ of $\hat \sigma_{es}$ allows for possible extra dephasing, while the decay rate $\gamma_\textrm{s}$ of $\hat \sigma_{gs}$ is due to dephasing only. In some experiments \cite{eisaman05}, $\gamma_\textrm{s}$ comes from the transverse diffusion of atoms out of the region defined by the quantized light mode. In these cases, the decay of $\hat \sigma_{gs}$ will be accompanied by population redistribution between states $|g\rangle$ and $|s\rangle$. In order to ensure that the corresponding incoming noise is vacuum (which our analysis requires, as we explain below and in Sec.~\ref{sec:effic}), we will assume in such cases that the incoming atoms are fully pumped into the level $|g\rangle$, which would correspond to a $2 \gamma_\textrm{s}$ decay rate of $\hat \sigma_{ss}$ into $\hat \sigma_{gg}$ (not included in Eqs.~(\ref{fullcavityeq}) since it does not affect the final equations). This is indeed the case if, as in Ref.~\cite{eisaman05}, the control beam diameter is much greater than the diameter of the quantized light mode. 

Assuming that almost all atoms are in the ground state at all times ($\hat \sigma_{gg} \approx N$ and $\hat \sigma_{ss} \approx \hat \sigma_{ee} \approx \hat \sigma_{es} \approx 0$), defining polarization $\hat P = \hat \sigma_{ge}/\sqrt{N}$ and spin wave $\hat S = \hat \sigma_{gs}/\sqrt{N}$, and working to first order in $\hat \eop$, we obtain Eqs.~(\ref{caveqs1})-(\ref{caveqs3}), where $\hat F_P = \hat F_{ge} /\sqrt{2 \gamma N}$ and $\hat F_S = \hat F_{gs}/\sqrt{2 \gamma_s N}$. 

Using the generalized Einstein relations \cite{cohen92, hald01}
\begin{eqnarray} \nonumber 
\langle \hat F_{\mu \nu}(t) \hat F_{\alpha \beta}(t') \rangle &=& \langle D(\hat \sigma_{\mu \nu} \hat \sigma_{\alpha \beta}) - D(\hat \sigma_{\mu \nu}) \hat \sigma_{\alpha \beta} 
\\ 
 &&- \hat \sigma_{\mu \nu} D(\hat \sigma_{\alpha \beta})\rangle \delta(t - t'),
\end{eqnarray}
where $D(\hat \sigma_{\mu \nu})$ denotes the deterministic part (i.e., with noise omitted) of the equation for $\dot{\hat{\sigma}}_{\mu \nu}$ in Eqs.~(\ref{fullcavityeq}), and again using the approximation that almost all atoms are in the ground state, we find that the only nonzero noise correlations between $\hat F_P$, $\hat F_S$, $\hat F_P^\dagger$, and $\hat F_S^\dagger$ are 
\begin{equation}
\langle \hat F_P(t) \hat F^\dagger_P(t') \rangle = \langle \hat F_S(t) \hat F^\dagger_S(t') \rangle = \delta(t-t').
\end{equation}
The fact that normally ordered correlations are zero means that the incoming noise is vacuum, which is precisely the reason why, as noted in Sec.~\ref{sec:effic}, efficiency is the only number we need in order to fully characterize the mapping. The property of Eqs.~(\ref{fullcavityeq}) that guarantees that the incoming noise is vacuum is the absence of decay out of state $|g\rangle$ into states $|e\rangle$ and $|s\rangle$. The decay into state $|e\rangle$ does not happen because the energy of an optical transition (on the order of $10^4$ K) is much greater than the temperature, at which typical experiments are done. In contrast, the energy of the $|s\rangle-|g\rangle$ transition in some experiments, such as the one in Ref.~\cite{eisaman05}, may be smaller than the temperature. However, the $|s\rangle-|g\rangle$ transition is typically not dipole allowed, so that the rate of $|g\rangle$ decay into $|s\rangle$ can be neglected, as well. As noted above, for the case when atoms are diffusing in and out of the quantized light mode, to keep the decay rate of $|g\rangle$ zero, we assume that the incoming atoms are fully pumped into $|g\rangle$. 

From Eq.~(\ref{cavcomrel}) and with the usual $\hat \sigma_{gg} \approx N$ assumption, we have
\begin{eqnarray}
\left[\hat S(t),\hat S^\dagger(t)\right] = 1,
\\
\left[\hat P(t),\hat P^\dagger(t)\right] = 1.
\end{eqnarray}
In particular, this means that $\hat S$ can be thought of as an annihilation operator for the spin-wave mode, into which we would like to store the state of the incoming photon mode.

The input and output fields, which propagate freely outside of the cavity, satisfy \cite{walls94}
\begin{equation} \label{cavecom}
\left[\hat \eop_\textrm{in}(t),\hat \eop^\dagger_\textrm{in}(t')\right] = \left[\hat \eop_\textrm{out}(t),\hat \eop^\dagger_\textrm{out}(t')\right] = \delta(t-t')
\end{equation} 
and can be expanded in terms of any orthonormal set of field (envelope) modes $\left\{h_\alpha(t) \right\}$ defined for $t \in [0,\infty)$, satisfying the orthonormality relation $\int_0^\infty d t h^*_\alpha(t) h_\beta(t) = \delta_{\alpha \beta}$ and completeness relation $\sum_\alpha h^*_\alpha(t) h_\alpha(t') = \delta(t-t')$, as
\begin{eqnarray}
\hat \eop_\textrm{in}(t) = \sum_\alpha h_\alpha(t) \hat a_\alpha,
\\
\hat \eop_\textrm{out}(t) = \sum_\alpha h_\alpha(t) \hat b_\alpha,
\end{eqnarray}
where annihilation operators $\left\{\hat a_\alpha \right\}$ and $\left\{\hat b_\alpha \right\}$ for the input and the output photon modes, respectively, satisfy
\begin{equation}
\left[\hat a_\alpha, \hat a^\dagger_\beta\right] = \left[\hat b_\alpha, \hat b^\dagger_\beta\right] = \delta_{\alpha \beta}.
\end{equation} 

Repeating for clarity the setup from Sec.~\ref{sec:cavmod}, we recall that all atoms are initially pumped into the ground state, i.e., no $\hat P$ or $\hat S$ excitations are present in the atoms. We also assume that the only input field excitations initially present are in the quantum field mode with annihilation operator $\hat a_0$ and envelope shape $h_0(t)$ nonzero on $[0,T]$. The goal is to store the state of this mode into $\hat S$ and at a time $T_\textrm{r} > T$ retrieve it back onto a field mode. During storage, we can, in principle, solve the operator Eqs.~(\ref{caveqs1})-(\ref{caveqs3}) for $\hat S(T)$ as some linear functional of $\hat \eop_\textrm{in}(t)$, $\hat F_P(t)$, $\hat F_S(t)$, $\hat S(0)$, and $\hat P(0)$. The storage efficiency is then given by
\begin{equation}
\eta_{\textrm{s}} \!=\! \frac{(\textrm{number of stored excitations})}{(\textrm{number of incoming photons})} \!=\! \frac{\langle \hat S^\dagger(T) \hat S(T) \rangle}{\int_{0}^T\!\!  d t  \langle \hat \eop^\dagger_\textrm{in}(t) \hat \eop_\textrm{in}(t) \rangle}.
\end{equation}
Since $\hat S(0)$ and $\hat P(0)$ give zero when acting on the initial state, and since all normally ordered noise correlations are zero, only the term in $\hat S(T)$ containing $\hat \eop_\textrm{in}(t)$ will contribute to the efficiency. Moreover, $h_0(t) \hat a_0$ is the only part of $\hat \eop_\textrm{in}(t)$ that does not give zero when acting on the initial state. Thus, for the purposes of finding the storage efficiency, we can ignore $\hat F_P$ and $\hat F_S$ in Eqs.~(\ref{caveqs1})-(\ref{caveqs3}) and treat these equations as complex number equations with $P(0) = 0$, $S(0) = 0$, and $\eop_\textrm{in}(t) = h_0(t)$. We have here dropped the carets on the operators to denote their complex number representations. To get back the nonvacuum part of the original operator from its complex number counterpart, we should just multiply the complex number version by $\hat a_0$.

Similarly, during retrieval, we can ignore $\hat F_P(t)$ and $\hat F_S(t)$ and can treat Eqs.~(\ref{caveqs1})-(\ref{caveqs3}) as complex number equations with the initial and boundary conditions given in Sec.~\ref{sec:cavmod}.

\section{Shaping the Control Field for the Optimal Adiabatic Storage \label{sec:appstor}}

In this appendix, we present the derivation of Eq.~(\ref{cavstoptom}), which gives the optimal storage control field during adiabatic storage. We then verify that this optimal control is just the time reverse of the control that retrieves into the time reverse of the input mode. 

To solve for the control field $\Omega(t)$ that maximizes the storage efficiency $\eta_\textrm{s}$ in Eq.~(\ref{cavetas}), we note that $f(t)$ defined in Eq.~(\ref{cavadf}) satisfies $\int_0^T |f(t)|^2 dt \leq 1$, with the equality achieved when 
\begin{equation} \label{cavsthcond}
\frac{2 \gamma (1+C)}{\gamma^2(1+C)^2+ \Delta^2} h(0,T) \gg 1, 
\end{equation}
which is equivalent to the requirement we had in Eq.~(\ref{cavrethcond}) for complete retrieval.
Since we also have $\int_0^T |\eop_\textrm{in}(t)|^2 = 1$, the integral in Eq.~(\ref{cavetas}) can be seen as a simple scalar product between states, and the efficiency is therefore $\eta_\textrm{s} \leq C/(1+C)$ with the equality achieved when (up to an undefined overall unimportant phase)
\begin{equation}\label{cavimped}
f(t) = \eop_\textrm{in}^*(t).
\end{equation}
We will now show that, for any given $\eop_\textrm{in}(t)$, $\Delta$, and $C$, there is a unique control that satisfies Eq.~(\ref{cavimped}) and thus gives the maximum storage efficiency $C/(1+C)$. In Refs.~\cite{yelin00, sherson05}, this control was found through a quantum impedance matching Bernoulli equation obtained by differentiating Eq.~(\ref{cavimped}). In order to be able in paper II to generalize more easily to free space,
we will solve Eq.~(\ref{cavimped}) directly. To do this, we follow a procedure very similar to that in Sec.~\ref{sec:cavadret}. We integrate the norm squared of Eq.~(\ref{cavimped}) from $0$ to $t$ to get
\begin{equation}\label{cavint}
\int_0^t |\eop_\textrm{in}(t')|^2 d t' = e^{-\frac{2 h(t,T)  \gamma (1+C)}{\gamma^2(1+C)^2+ \Delta^2}}-e^{-\frac{2 h(0,T) \gamma (1+C)}{\gamma^2(1+C)^2+ \Delta^2}}.
\end{equation}
Since $h(T,T) = 0$, the normalization of $\eop_\textrm{in}(t)$ requires the satisfaction of Eq.~(\ref{cavsthcond}). Assuming it is satisfied to the desired precision, we solve Eq.~(\ref{cavint}) for $h(t,T)$, and then taking the square root of the negative of its derivative with respect to $t$, we find $|\Omega(t)|$. Knowing $h(t,T)$, the phase of $\Omega(t)$ can then be determined from Eq.~(\ref{cavimped}). Putting the magnitude and the phase together, we obtain the expression for the optimal control given in Eq.~(\ref{cavstoptom}).

We will now show that, as expected from the time-reversal argument, the optimal control we derived is just the time reverse ($\Omega(t) \rightarrow \Omega^*(T-t)$) of the control that retrieves into $\eop^*_\textrm{in}(T-t)$, the time reverse of the input mode. To see this, we note that from Eq.~(\ref{eeq}) it follows that the magnitude of the control field $\Omega(t)$ that retrieves into $e(t) = \eop^*_\textrm{in}(T-t)$ is determined by
\begin{equation} \label{heq}
\int_0^t d t' |\eop_\textrm{in}(t')|^2 = e^{-\frac{2 \gamma (1+C)}{\gamma^2 (1+C)^2 + \Delta^2} h(0,T-t)}.
\end{equation}
Putting $e(t) = \eop^*_\textrm{in}(T-t)$ into Eq.~(\ref{cavretoptom}), taking the complex conjugate of the result, and evaluating at $T-t$, we get 
\begin{equation} \label{omstar}
\Omega^*(T\!-t)\! =\! - \frac{\gamma(1+C)\!-\! i \Delta}{\sqrt{2 \gamma (1+C)}} \frac{\eop_\textrm{in}(t)}{\sqrt{\!\int_0^t  |\eop_\textrm{in}(t')|^2 d t'}} e^{i \frac{\Delta h(0,T-t)}{\gamma^2 (1+C)^2 + \Delta^2} }.
\end{equation}
Since $h(0,T-t)$ determined by Eq.~(\ref{heq}) and $h(t,T)$ determined by Eq.~(\ref{cavint}) are equal, the right-hand side of Eq.~(\ref{omstar}) is, in fact, equal to the right-hand side of Eq.~(\ref{cavstoptom}), as desired.


\begin{thebibliography}{99}
\bibitem{polziknote} We note for the sake of clarity that the storage scheme employed in Ref.~\cite{polzik04} cannot be regarded as a subset of the general scheme considered in the present work, and more work has to be done in order to compare the performance of the two.
\bibitem{vuleticnote} For clarity, we note that, although it might, in principle, be possible to describe the experimental setup in Ref.~\cite{vuletic06} by our model, this setup is slightly different from our model in that the propagation directions of the control and the photon are orthogonal (rather than the same) and the cavity mode is a standing-wave mode (rather than a running-wave mode).
\bibitem{complexconjugate} To be more precise, such time-reversed storage stores into the complex conjugate of the original spin-wave mode. We study in paper II how, for the case of nonzero energy difference $\omega_{sg}$ between states $|g\rangle$ and $|s\rangle$, the position-dependent phase $\exp(2 i z \omega_{sg}/c )$, by which a given spin wave and its complex conjugate differ, reduces the optimal efficiency of storage followed by backward retrieval, where backward retrieval means that the retrieval and storage controls are counter-propagating. However, in the present paper, we assume that, for storage followed by retrieval, the $k$ vectors of the storage and retrieval control fields are pointing in the same direction, in which case nonzero $\omega_{sg}$ is not a problem.
\bibitem{cavinitialcondition} To be precise, another condition in addition to Eq.~(\ref{cavPdot}) required for the validity of the adiabatic elimination is that $P(0)$ derived from adiabatic elimination must be much less than $1$, so that it is close to the given value of $P(0) = 0$. During retrieval, this condition, however, is equivalent to condition (\ref{cavcond1}) evaluated at $t = 0$. Similarly, during storage of a sufficiently smooth pulse (which will, thus, satisfy $\eop_\textrm{in}(0) \lesssim 1/\sqrt{T}$), the desired condition on the adiabatic value of $P(0)$ is implied by condition (\ref{cavcond}). Thus, we do not state this condition on the adiabatic value of $P(0)$ separately.
\end{thebibliography}
\end{document}